\documentclass[prl,byrevtex,balancelastpage,letterpaper,10pt,showpacs,preprint]{revtex4}
\usepackage{amssymb}
\usepackage{amsmath}
\usepackage{graphicx}

\newcommand {\be}{\begin{equation}}
\newcommand {\ee}{\end{equation}}
\newcommand {\bea}{\begin{eqnarray}}
\newcommand {\eea}{\end{eqnarray}}

\begin{document}

\title{Cavity cooling of an optically trapped nanoparticle}

\author{P. F. Barker}
\affiliation{Department of Physics and Astronomy, University College London, WC1E 6BT, United Kingdom}
\author{M. N. Shneider}
\affiliation{Applied Physics Group, Department of Mechanical and Aerospace Engineering, Princeton University, Princeton, NJ 08544}

\begin{abstract}
 We study the cooling of a dielectric nanoscale particle trapped in an optical cavity. We derive the frictional force for
 motion in the cavity field, and show that the cooling rate is proportional to the square of oscillation amplitude and frequency. Both the radial and axial centre-of-mass motion of the trapped particle, which are coupled by the cavity field, are cooled. This motion is analogous to two coupled but damped pendulums. Our simulations show that the nanosphere can be cooled
 to e$^{-1}$ of its initial momentum over time scales of hundredths of milliseconds.
\end{abstract}

\pacs{37.10.Vz,  37.30.+i} \maketitle \preprint{} \eid{} \startpage{1}
\endpage{}

The study of micro- and nanomechanical oscillators \cite{nanoosc1,nanoosc2,squid, karrai}, and particularly their quantum mechanical motion \cite{braginskii, quantumlims, penrose}, is a rapidly developing field which promises insights into the boundary between quantum and classical worlds. Also, their sensitivity to the environment appears promising for the development of chemical sensors with single atom sensitivity \cite{nanomass}.  Cavity opto-mechanics is an important area within this field, in which the mechanical motion of at least one degree of the freedom of an oscillator is damped or cooled by interaction with the field of an optical cavity \cite{reviewopto, karrai, resolvedsideband,membrane, kippenberg}. Cooling of the mechanical motion works by coupling an optical cavity to the oscillator so that it selectively scatters blue shifted photons of a probe beam out of the cavity with respect to the incident photons.  By conservation of energy, the mechanical energy of the oscillator-cavity system is reduced.  A range of optomechanical oscillators and cooling schemes have been now been realized experimentally \cite{reviewopto}, and currently there is a strong impetus towards reaching the quantum regime, where only a few of the quantised states of at least one degree of freedom are occupied. The cooling of atomic species using laser cooling is now well established with the creation of nanoKelvin temperature gases, which has led to the realization of quantum degeneracy in gases. For molecular and atomic species that cannot be laser cooled, cavity cooling of a trapped species appears attractive because it does not rely on the detailed internal level structure \cite{horak, vuleticcooling}. This scheme has already been used to cool a trapped atom \cite{rempe}, an ion \cite{ion}, and atomic gases \cite{vuletic1}.

\begin{figure}[h]
\includegraphics[scale=0.4]{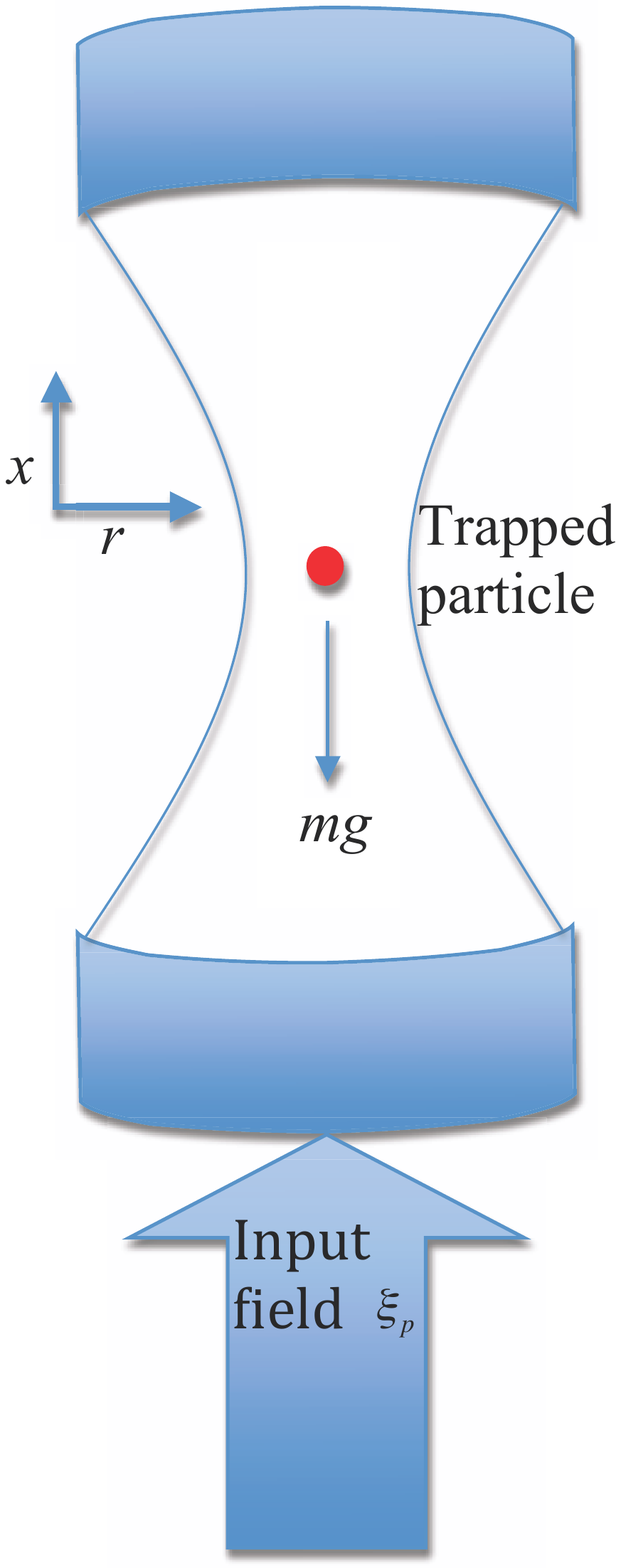}
\caption{Schematic diagram for cooling a small dielectric nanoscale particle in a high finesse cavity.} \label{fig1}
\end{figure}

\section{I. introduction}

In this paper we consider the cavity cooling of a nanoscale particle trapped at the center of a high finesse Fabry-Perot cavity in a vacuum.  Here a center-of-mass oscillation occurs due to trapping in the periodic potential of the interference pattern created inside the cavity, while cooling occurs by interaction with the field mediated by the cavity.  Unlike many opto-mechanical cooling schemes that utilize radiation pressure, this scheme uses the dipole force.  Cavity cooling in this way is attractive because a trapped particle can be effectively isolated from the environment. This is unlike the cooling of many cavity optomechanical schemes such as cantilever \cite{karrai} or membranes \cite{harris}, which are directly physically connected to a large heat bath.

We consider cavity cooling of a large polarizable particle that is trapped in the intra-cavity field of a high finesse cavity by modifying the model developed for a single atom in a 1-D cavity \cite{horak1}. Because we only consider a single trapped particle, we are able to derive a velocity dependent frictional force for small oscillations around the antinode of the intra-cavity field.  We then consider cooling in a realistic 2-D cavity which includes the damping of the axial and radial motion. We show that both degrees of freedom are coupled by the cavity field which acts to damp them.  In addition to dipole force due to the cavity field, which acts to trap the particle, the nanoparticle is also subject to gravity, which is chosen to be along the cavity axis as shown in figure 1.

\section{II. 1-D cooling of a nanosphere}

To explore cavity cooling we assume a field of amplitude $\xi_{ext}$ and frequency $\omega_p$ incident on one of the two high finesse cavity mirrors of equal reflectivity, $R$, with conductivity $\sigma$. The line width of the cavity is $\kappa$ and the 1/e lifetime of a photon in the cavity is $\frac{2\pi}{\kappa}$. We assume that the light is impedance matched to the cavity, that the cavity is stabilized such that only the TEM$_{00}$ mode propagates, and that the nanoparticle is trapped at its beam waist. We first determine the field within the cavity in the presence of the single dielectric particle from the one-dimensional wave equation given by
\bea \frac{\partial^2 E(x,t)}{\partial t}+ 2 \kappa\frac{\partial E(x,t)}{\partial t} - c^2 \frac{\partial^2 E(x,t)}{\partial x^2}=\nonumber \\
\frac{1}{\epsilon_0}\frac{\partial^2 P(x,t)}{\partial t^2}+2\kappa^{ext}\frac{\partial E(x,t)^{ext}}{\partial t}\eea
where $\kappa=\frac{\sigma}{2\epsilon_0}$ and $\kappa^{ext}=\frac{\sigma}{2\epsilon_0}^{ext}$.

Here $E(x,t)$ and $P(x,t)$ represent the sum of all possible allowed electric field and polarization modes for the cavity respectively. We assume that the cavity only operates on one of these modes where the electric field is $E_m(x,t)=\xi(t)(e^{-i\omega_{p}t}+c.c)cos(kx)$ and the polarization is given by $P_m=p(t)e^{-i\omega_{p}t}+c.c.$ in the cavity. The external field which couples to this cavity mode is given by $E_m^{ext}=\xi^{ext}(t)e^{-i\omega_{p}t}+c.c.$.  We find the equation for the evolution of field amplitude of this mode by utilizing the orthogonality of cavity modes and multiplying equation (1) by $E_m(x,t)$ and averaging over the cavity volume $V$.

Under the slowly varying envelope approximation, we obtain the 1-D equation of motion for the amplitude of the single mode field as

\be \frac{\partial \xi}{\partial t}=-[\kappa-i(U(x)+\Delta)]\xi+\kappa_{ext}\xi_{ext} \label{amplitude} \ee

where $U(x)=\frac{\alpha \omega_{p}\cos^2 kx}{\epsilon_0 V}$ is the position-dependent shift in cavity frequency induced by the polarizable particle and $\Delta = \omega_p-\omega_c$ is the cavity detuning from resonance.

The equation of motion for the momentum is simply the dipole force that results from the intra-cavity field acting against gravity

\be \frac{d P_{x}}{d t} = -\alpha |\xi|^2 k \sin{2 k x}-mg \label{momentum} \ee

and the particle position is given by

\be \frac{d x}{d t} =  P_{x}/m \label{position} \ee

It is not clear from the coupled equations \eqref{amplitude}, \eqref{momentum} and \eqref{position}
that there is a dissipative force on the trapped particle in the intracavity field. To understand the damping or heating of the trapped nanoparticle we derive an equation of motion with a velocity dependent frictional force.   As we are only interested in trapped motion, we need only consider small amplitude oscillations when $kx<<1$. Additionally, because the center-of-mass motion is much smaller than the round trip time for light in the cavity, we average equations \eqref{amplitude}, \eqref{momentum} and \eqref{position} over a time $\Delta t$ from $t-\Delta t$ to $t$. A suitable $\Delta t$ is the cavity decay time $2 \pi/ \kappa$. The averaged field $\overline{\xi}$ over $2 \pi/\kappa$ is given by

\be \overline{\xi }=\frac{\kappa^{ext}\xi^{ext}}{\kappa-i[\Delta+\frac{\alpha\omega_p}{\epsilon_0 V}+\frac{\alpha\omega_p k^2}{\epsilon_0 V}( v \frac{2 \pi x}{\kappa} - x^2)]}\ee

where the position dependent expression for the frequency shift $U(x)$ is expanded around $x =0$ so that $U(x)\approx\frac{\alpha \omega_{p} (1-(kx)^2)}{\epsilon_0 V}$ and $x(t-\Delta t)=x(t)-2 v\pi/ \kappa x(t)$, where $v=\frac{dx}{dt}$.

This enables us to find a time-averaged equation of motion for momentum.  This is given by $\overline{\frac{d p_{x}}{d t}}=M \frac{d^2 x}{d t^2}\approx -2\alpha k^2 \overline{\xi}^2 x$, which leads to an equation of motion of the trapped nanosphere which is a type of Li\'{e}nard differential equation \cite{lienard}. This is given by
\be
\frac{d^2 x}{dt^2}=-\Omega^2(x + \beta x^3-\frac{2 \pi \beta}{\kappa} x^2  v)  - mg\label{eqmotion}
\ee
where $\Omega=\sqrt{\frac{2\alpha \delta k^2}{M}}$, $\delta=\frac{|\kappa^{ext}\xi^{ext}|^2}{\kappa^2+(\Delta+\frac{\alpha \omega_p}{ \epsilon_0V})^2}$ and $\beta=\frac{\alpha \omega_p k^2 (\Delta + \frac{\alpha \omega_p}{ \epsilon_0 V})}{ \epsilon_0 V [\kappa^2 +(\Delta + \frac{\alpha \omega_p}{ \epsilon_0 V})^2]}$. This equation differs from conventional oscillators such as the van der Pol or Duffing type in that the friction term, $\frac{\beta}{\kappa} x^2$, is position dependent. When $\beta<0$  the oscillatory motion of the sphere is damped and therefore the cavity detuning is $\Delta<-\alpha \omega_p/ \epsilon_0 V$. Optimal damping or cooling of the sphere occurs when
\be \Delta_{op}=-\alpha \omega_p/\epsilon_0 V - \kappa. \label{optimal} \ee
An approximate expression for the energy loss rate averaged over an oscillation period $2\pi/\Omega$ can also be determined. The rate of change in the total energy, $E$, of the particle is $\frac{dE}{dt} = M v\frac{d v}{d t} + M \Omega^2 x \frac{dx}{dt} + M\beta \Omega^2 x^3 \frac{dx}{dt}$, while the power can be determined by multiplying the equation of motion (\ref{eqmotion}) by $Mv$ to give $Mv \frac{d^2 x}{dt^2}=-M\Omega^2 xv - M \Omega^2 \beta x^3 v+\frac{2 \pi \beta}{\kappa}M\Omega^2v^2x^2$. Equating equal terms gives the rate of change in total energy of the system as $\frac{dE}{dt}= \frac{2 \pi \beta}{\kappa}Mx^2 \Omega^2 v^2$. When averaged over an oscillation period, and assuming the damping is slow enough not to change the amplitude of oscillations during a cycle, we determine a time averaged exponential damping rate of
\be
\Gamma = \frac{\pi \beta}{2 \kappa}\Omega^2 x_0^2, \label{damprate}
\ee
where $x_0$ is the amplitude of the oscillation. Because the amplitude will decrease as the particle cools at a fixed input intensity, so the damping rate also decreases, as has previously been noted \cite{lu}.  However, the cooling rate is proportional to the square of the oscillation frequency.

To study the cooling of nanoscale polarizable particles, we consider a SiO$_2$ nanosphere of radius $r = 100$ nm and refractive index of n=1.45 trapped in an antinode of the intra-cavity field. A sphere of this size within the Rayleigh regime when $r< < \lambda$ and can be treated as a dipole scatterer with polarizability $\alpha =4 \pi \epsilon_0 \frac{n^2-1}{n^2+2}r^3$, determined to be $2.98\times10^{-32}$ Cm$^2$ V$^{-1}$.   Its mass, determined from its density ($\rho=2198$ Kg m$^{-3}$), is $9.2\times10^{-18}$ kg. We consider a cavity of length $L=10$ cm, with mirrors of reflectivity R = 0.99995 resulting in $\kappa = 4.71\times10^{5}$ rad/s.  The intracavity field has a wavelength of 1064 nm with an input intensity of 100 mW and a beam radius of 50 microns in the cavity. The axial oscillation frequency for these parameters is $1.55\times10^{6}$ rad/s. The optimal detuning of the cavity from resonance is calculated from the above parameters to be $-5.13\times 10^{5}$ rad/s for cooling and $4.29 \times10^5$ rad/s for heating.  It is instructive to plot the energy damping/heating rate as a function of both detuning and position, as shown in figure 2. Peak negative values correspond to maximal cooling and peak positive values for maximal heating.  Figure 2 also shows that the damping/heating is larger when the particle is further from the anti-node of the interference pattern of the intracavity field, which has important consequences for cooling because it implies that the best damping will occur when the particle is not strongly confined and will undergo larger oscillations.

\begin{figure}[h]
\includegraphics[scale=0.5]{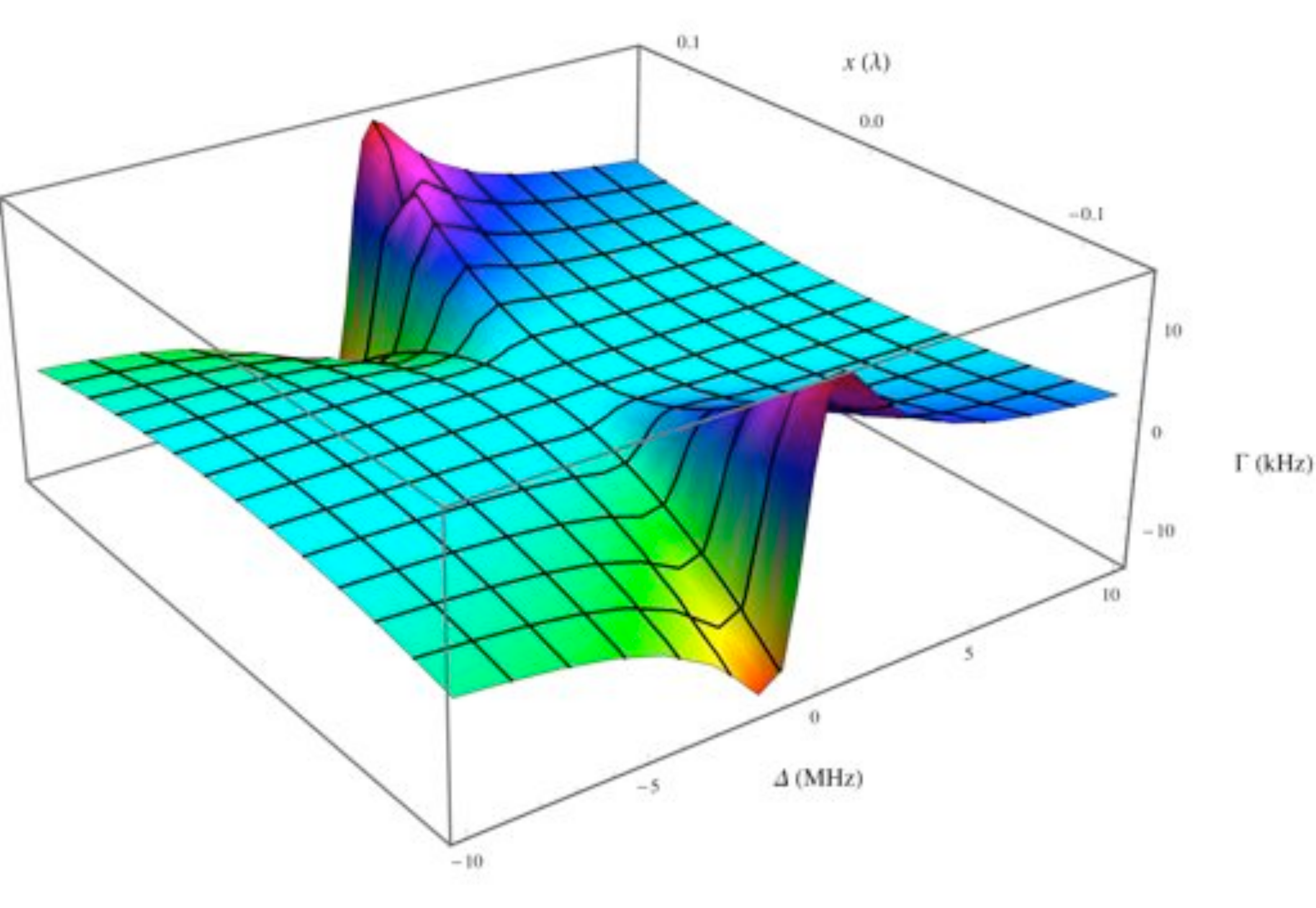}
\caption{A 3-D plot showing the variation of energy damping rate as a function of cavity detuning $ \Delta$ and oscillation amplitude of nanosphere position $x_0$. Maximal cooling occurs at negative detuning.} \label{frictionplot}
\end{figure}

%\begin{figure}[h]
%\includegraphics[scale=0.8]{timeseries}
%\caption{A plot of the velocity of the nanosphere over 1 ms indicating the damped oscillatory motion of the the particle trapped in the cavity %field.  The nanosphere has an initial velocity of 0.5 m/s at $x=0$.} \label{timeseries}
%\end{figure}

\begin{figure}[h]
\includegraphics[scale=1]{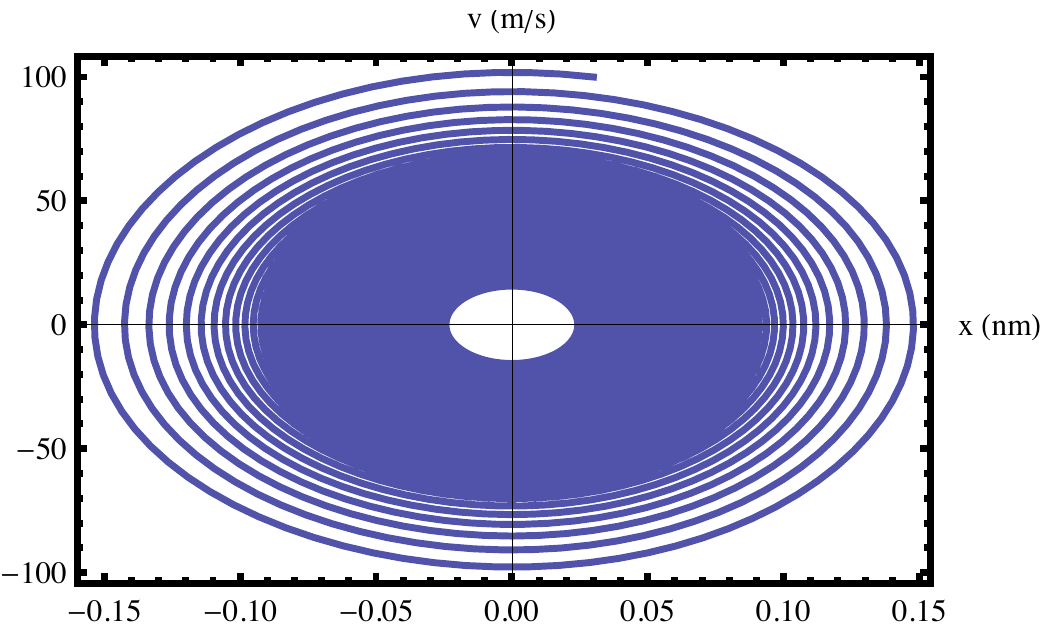}
\caption{A phase space plot showing damped motion in position over 1 ms. Note that the cooling rate decreases as the particle becomes localized in the lattice potential.} \label{phaseplot}
\end{figure}

To illustrate the damped oscillatory motion of the trapped nanosphere, we plot its velocity and position for an initial velocity of $v=3$ cm s$^{-1}$, corresponding to an initial kinetic energy of $E/k_b = 300 K$ and $x=100$ nm from trap center with optimal cavity detuning for damping. The equations of motion were integrated in time using a fourth-order Runge-Kutta method. The damped oscillatory motion is clearly shown, even for this short  time period $t=1 $ms, and there is good agreement between the simplified time-averaged equation of motion (\ref{eqmotion}) and the exact model of equations (\ref{amplitude},\ref{momentum},\ref{position}).   The oscillation frequency, largely determined by the input intensity, has been chosen to be smaller than the cavity line width cavity.  However, an increase in intensity and thus oscillation frequency can be used to cool the particle in the resolved-sideband limit, of importance for cooling to the quantum ground state \cite{resolved}.  Figure 3 shows a phase space plot of the velocity and position trajectory of the particle trapped in a single anti-node. The inward spiral indicates damping with a much tighter spiral as the particle approaches  $x = 0$, indicating a weaker frictional force as the particle approaches the anti-node.
It has been shown previously through simulations that atoms can be cooled more rapidly in cavity cooling schemes if the laser input intensity is lowered as the particle is cooled \cite{lu}. This can now be seen to be due to the position dependence of the damping rate. For high intensities the particle is tightly confined and the amplitude of the oscillations are small resulting in a lower cooling rate. To illustrate the differences in cooling time for a constant intensity field, we plot the amplitude of the oscillating kinetic energy of the particle as a function of time for a constant intensity of 1 mW input into the cavity and for an exponentially decaying input field, initially at 1 mW, with a time constant of 50 ms. We note that the lowest intensity that can be used is limited by gravity since eventually the well depth of the intra-cavity field becomes less than the gravitational potential. This occurs in figure 4 at approximately 0.35 s. Here, the cavity intensity must be held constant and the cooling rate where the cooling rate will be slower.

\begin{figure}[h]
\includegraphics[scale=0.5]{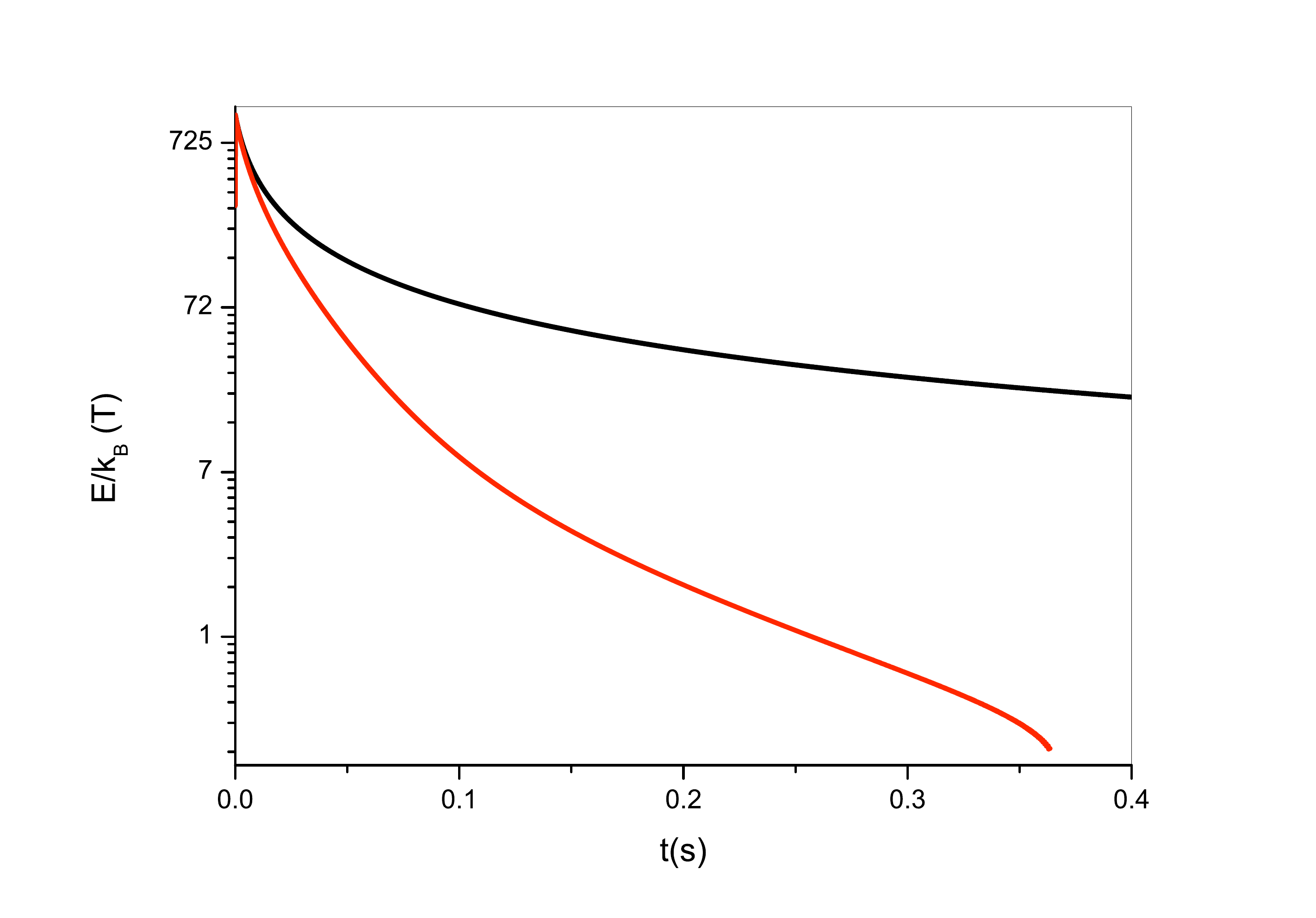}
\caption{Plot of the amplitude of the oscillatory energy as a function of time to compare of cooling times for a constant input intensity into the cavity of 1 mW (blue line) and an exponentially decreasing input intensity with e$^{-1}$ decay constant of 50 ms (red line). } \label{phaseplot}
\end{figure}

\section{III. Cooling in 2-D}
We now consider the experimentally more realistic situation of cooling in a cavity which has field variations in the axial and in the radial direction. We now must also consider motion in the radial direction as well.
 For this case we assume the field input into the cavity is that of the lowest order transverse electromagnetic mode
\be
E(r,x,t)=A(t)\frac{\omega_0}{\omega(x)}e^{-\frac{r^2}{\omega(x)^2}}e^{-i(\frac{kr^2}{2R(x)}+k x-\eta)}
\ee
where $\omega_0=\sqrt{L/k}$, $R(x)=\frac{1}{x}(x^2+x_0^2)$, $\eta=\arctan{\frac{x}{x_0}}$ and $\omega(x)=\omega_0\sqrt{1+(\frac{x}{x_0})^2}$. We require that the nanosphere is trapped at the center of the cavity where the intra-cavity intensity is highest. At this position ($x_0=L/2$) $\omega(x)\approx \sqrt{2}\omega_{0}$, $R(x)\approx 2 x_0$ and $\eta \approx \frac{\pi}{4}$.
 It is useful to recast this in trigonometric form as
 $E(r,x,t)= \frac{\xi}{2}\exp(-\frac{r^2}{\Lambda^2})[\cos
  kx(\cos\frac{r^2}{\Lambda^2}+\sin{\frac{r^2}{\Lambda^2}})-\sin kx (\sin \frac{r^2}{\Lambda^2}-\cos \frac{r^2}{\Lambda^2})]e^{-i\omega_{p}t}+ c.c.$, where $\Lambda=(2L/k)^{1/2}$

The temporal evolution of the field amplitude within a cavity is then determined from the 2-D wave equation under the slowly varying envelope approximation as for the 1-D case above. This is now given by

\be \frac{\partial \xi}{\partial t}=-[\kappa-i(U(x)+\Delta)]\xi+\kappa_{ext}\frac{\Lambda_{ext}^2}{\Lambda^2}\xi_{ext}, \ee

All other variables are the same as for the 1-D case except that $U(x)=\frac{\alpha \omega_p F(x,r)}{\pi \epsilon_0 \Lambda^2 L}$ and $F(x,r)=e^{-\frac{r^2}{\Lambda^2}}[\cos^2{k x}\,(\cos{\frac{r^2}{\Lambda^2}}+\sin{\frac{r^2}{\Lambda^2}})-\frac{1}{2}\sin{2kx}\,(\cos{\frac{r^2}{\Lambda^2}}-\sin{\frac{r^2}{\Lambda^2}})]$

The rate of change of the momentum and position in the axial direction and radial direction within the cavity is given as

\bea \frac{d P_{x}}{d t} =
-\alpha k |\xi|^2 e^{-\frac{2r^2}{\Lambda^2}}[\sin{(2 k x)}\, \sin{\frac{2r^2}{\Lambda^2}}-
\nonumber \\ \cos{(2kx)}\, \cos{\frac{2 r^2}{\Lambda^2}}] - mg \eea

\bea \frac{d P_{r}}{d t} = -2\alpha |\xi|^2 \frac{r^2}{\Lambda^2} e^{-\frac{2r^2}{\Lambda^2}}\, [1+\cos{(2kx)}(\sin{ \frac{2r^2}{\Lambda^2}}-\nonumber \\\cos{\frac{2r^2}{\Lambda^2}})+\sin{(2kx)}(\sin{\frac{2 r^2}{\Lambda}}+ \cos{\frac{2 r^2}{\Lambda^2}})]
\eea

\be \frac{d x}{d t} =  P_{x}/m \ee

\be \frac{d r}{d t} =  P_{r}/m \ee

These equations describe the axial and radial oscillations of the nanosphere coupled by the intra-cavity field, which also acts to damp the motion in both directions. We integrate these equations in time using the Runga-Kutta scheme for a nanosphere ($r$ = 100 nm) trapped in a cavity operating in the fundamental TEM$_{00}$ mode. A optical field of wavelength 1064 nm and power of 1 mW is input into the cavity with a cavity spot size of $\Lambda$=58 $\mu$m. The cavity has $\kappa=4.7\times10^{6}$ rad/s and a length of 1 cm. Figure 5 is a plot of the axial and radial velocity of the nanosphere as a function of time with initial conditions of $x=100$ nm , $r =15$ $\mu$ m and $v_x=v_r= 3$ cm/s.  Oscillation frequencies of 476 Hz (radial) and 122 kHz (axial) result from this fixed intracavity intensity ($3.8\times10^9$ Wm$^{-2}$) and cavity geometry. The very different frequencies result from the large difference in the axial and radial field gradients at the center of the cavity. Note that in this figure the fast axial motion is modulated by the slower radial motion via the cavity field.  This small modulation of the cavity field by this motion is shown in figure 6. The cavity detuning for these simulations was chosen to be $\Delta_c=-\alpha\omega_p/\epsilon_0V - \kappa$ optimized for cooling of the axial motion. The radial motion could be more effectively cooled by a smaller detuning.

\begin{figure}[h]
\includegraphics[scale=0.5]{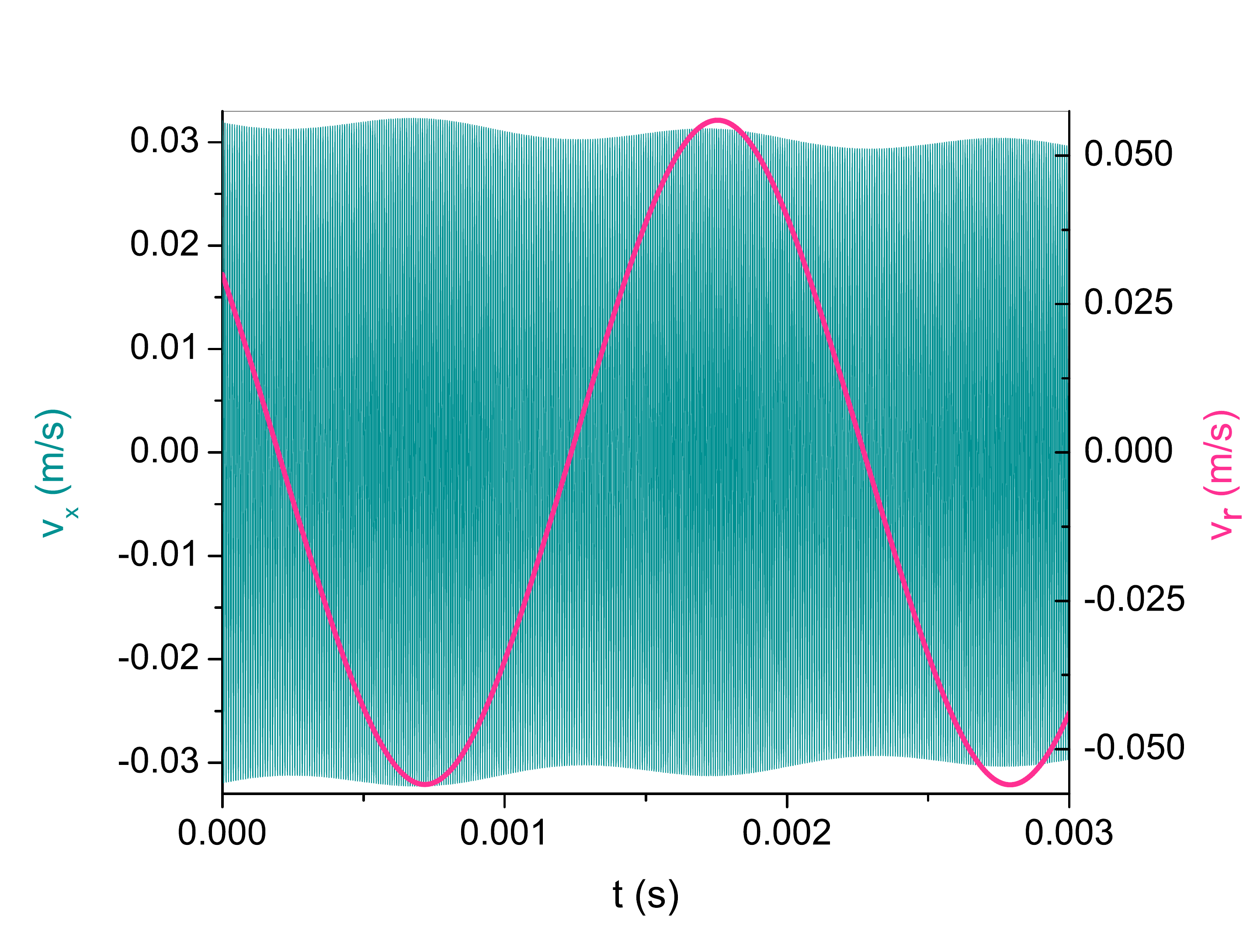}
\caption{Plots illustrating the difference in oscillation frequency of the axial and radial velocity of the trapped nanosphere. The reduction in the amplitude of the axial oscillation indicates cooling over this small time period.} \label{phaseplot}
\end{figure}

\begin{figure}[h]
\includegraphics[scale=0.5]{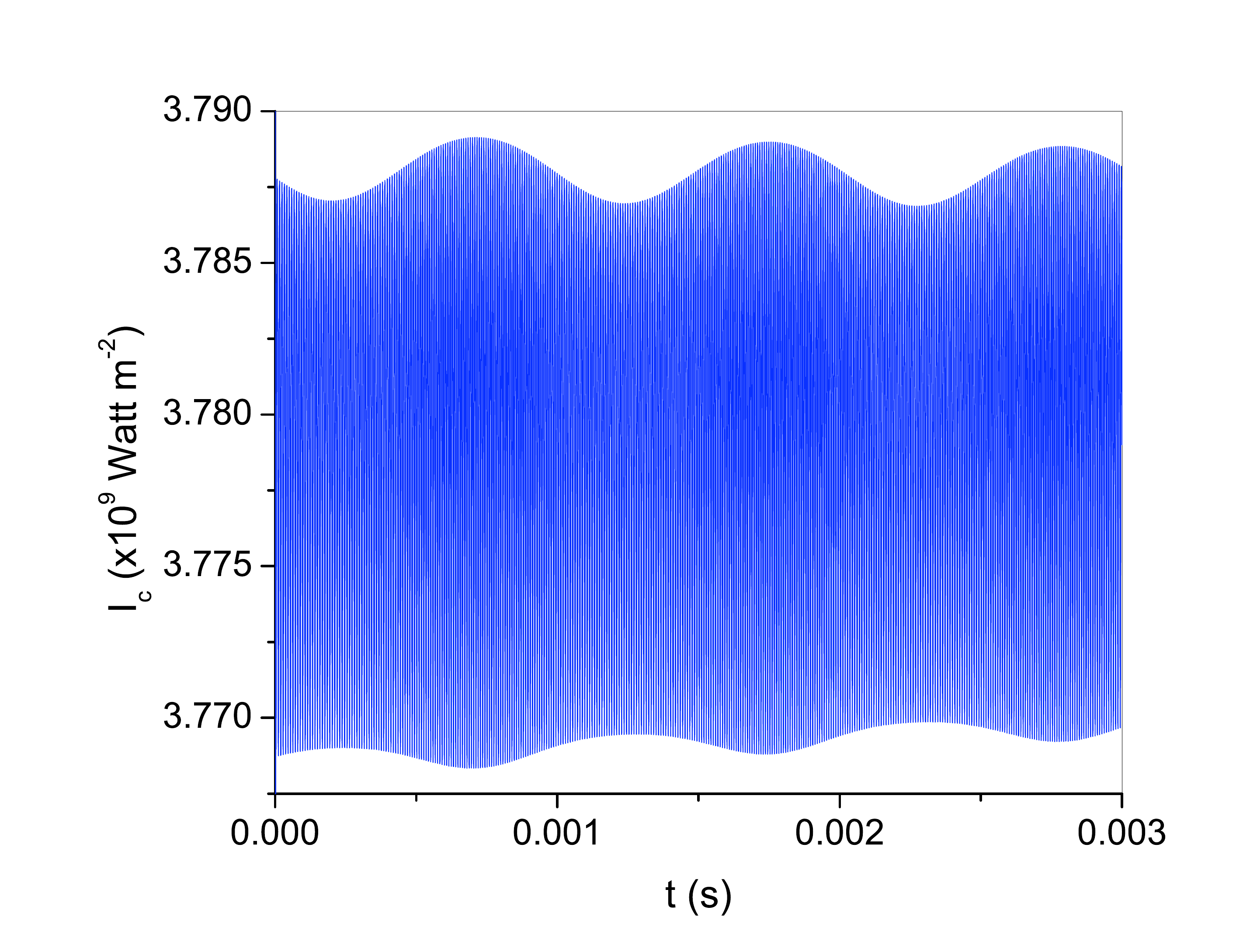}
\caption{The oscillation in the cavity intensity during the first 3 ms of cooling where modulation of the intensity by the radial and axial motion can be seen when compared to the axial and radial velocity shown in figure 6. } \label{phaseplot}
\end{figure}

\begin{figure}[h]
\includegraphics[scale=0.5]{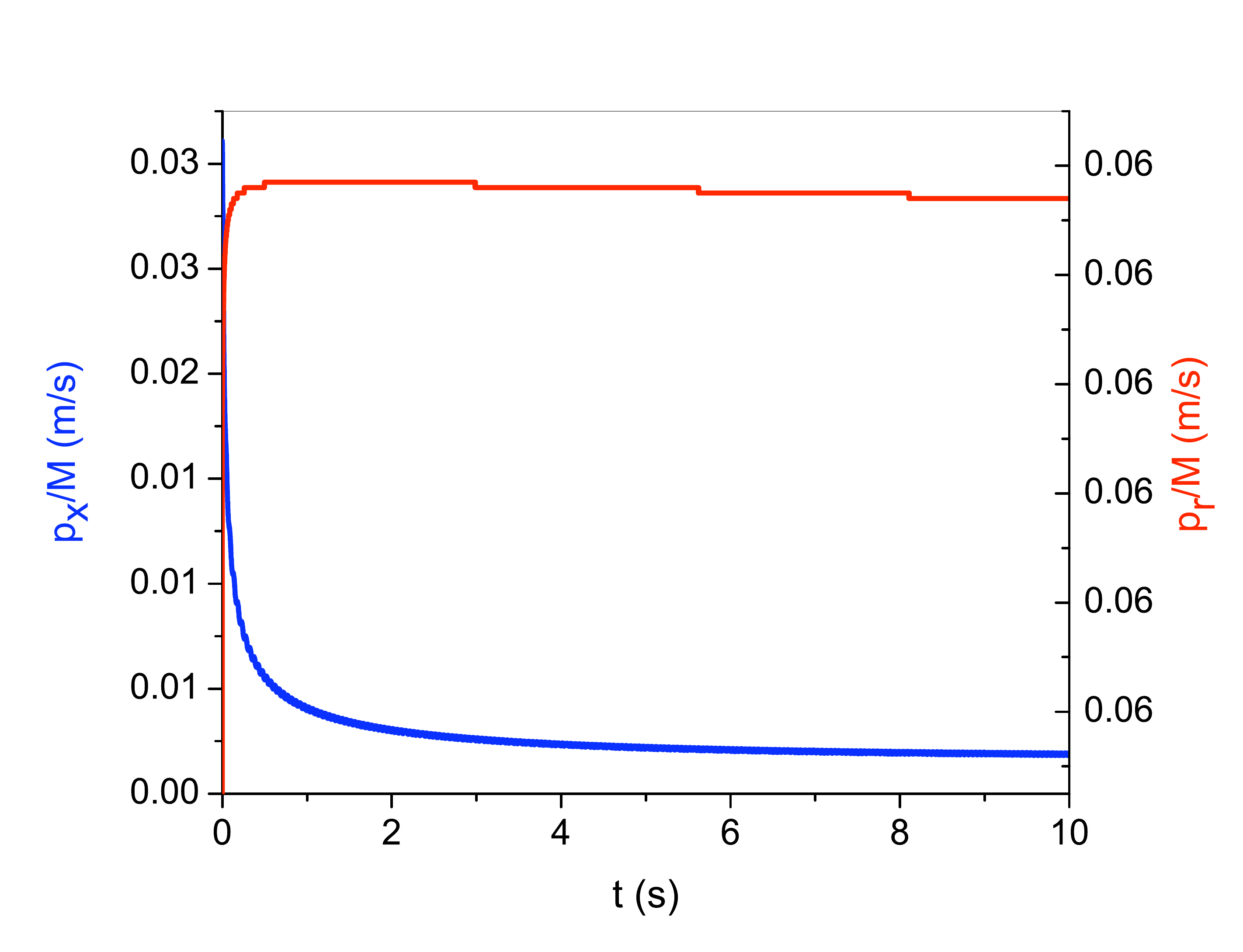}
\caption{A plot of the amplitude of the oscillatory velocity of the particle for both the axial ($v_{x}$) and radial ($v_{r}$) oscillatory motion as a function of time for an input field of constant intensity. The blue curve is the axial motion and the red curve is the radial velocity.}\label{phaseplot1}
\end{figure}

Figure 7 is a plot of the absolute value of the envelope of the oscillatory velocity of each degree of freedom as a function time. Note that the radial cooling is much slower that in the axial direction. This is as expected from the simple 1-D analysis which shows that the damping rate is proportional to the square of the oscillation frequency. Here the radial frequency is more than two orders of magnitude less than the axial frequency. To verify this we have artificially increased the radial gradient while keeping the intensity constant so that the radial frequency is the same as the axial frequency. In this case both degrees of freedom cool at approximately the same rate. It is well known that coupled oscillators or pendulums typically establish a constant phase difference depending on whether the coupling is damped or undamped. We have simulated the cooling of the nanosphere for both a fixed input intensity and with a decaying intensity with a time constant of 50 ms.
The cooling of both degrees of freedom is shown in a plot of the total kinetic energy as a function of time in figure 8. This figure clearly demonstrates that the nanosphere can be cooled with a decaying input intensity and constant input intensity. As before, the cooling rate is faster with a decaying intracavity intensity, until the particle can no longer be trapped by the field, at t = 0.2 s in this case. An external potential, which supports the particle against gravity, would allow more rapid cooling to lower temperatures. This could be another optical field that is not resonant with the cavity field or a nanoscale charged particle within an ion trap \cite{ashkin,schlemmer}. The cooling of the particle can be monitored by observing scattered light or the light transmitted through the cavity. A Fourier transform of the light will show the motional sidebands for both degrees of freedom, whose amplitude will decrease as the particle cools.

\begin{figure}[h]
\includegraphics[scale=0.5]{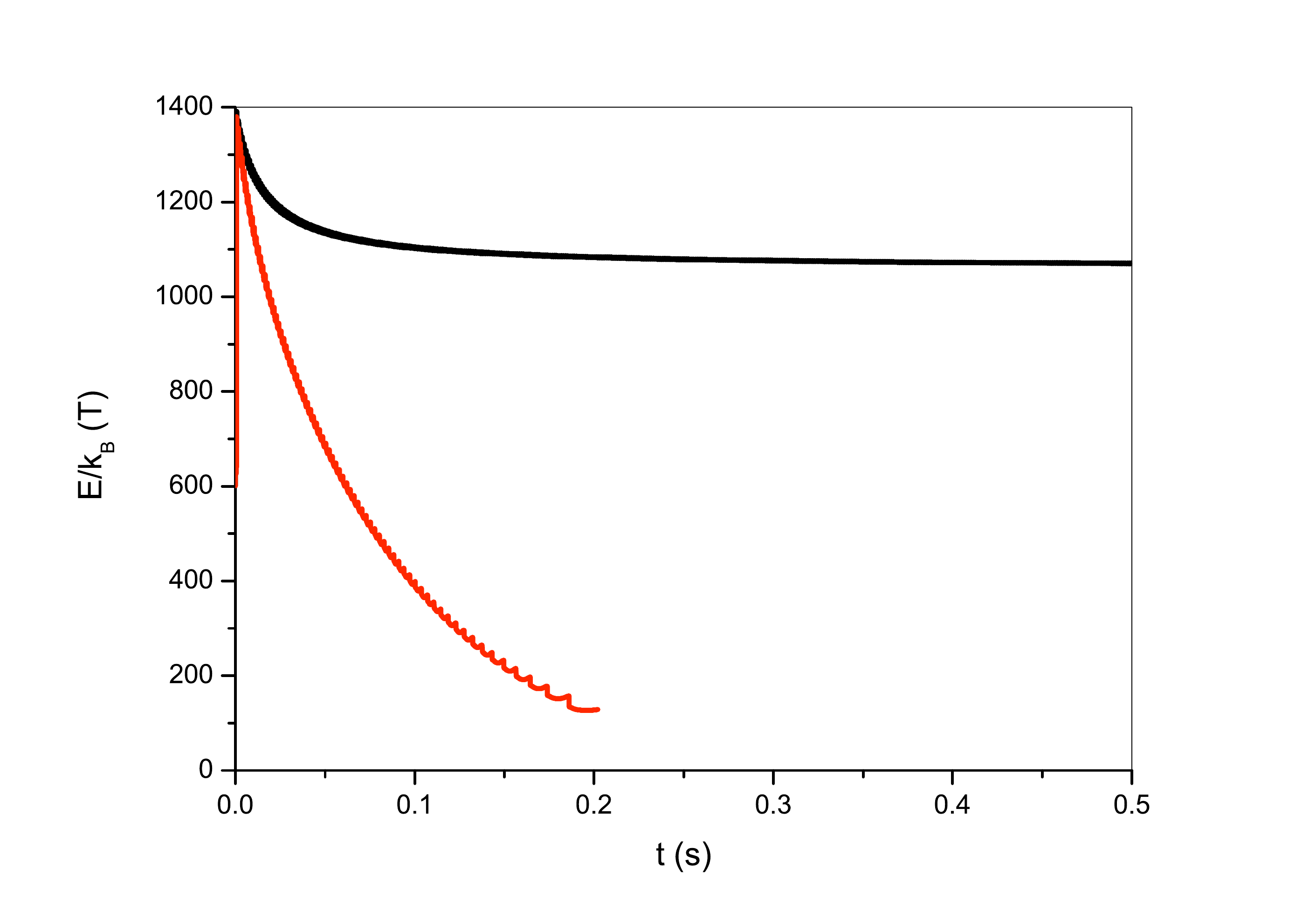}
\caption{Plots of the decay of the amplitude of oscillating total kinetic energy in the cavity as a function of time, including both axial and radial motion.  The blue curve is for fixed input intensity and the red is for an exponential decaying input intensity with a time constant of 50 ms.} \label{phaseplot}
\end{figure}

\section{IV. Conclusions}

We show that cooling nanoscale dielectric particles is feasible in the classical regime using cavity cooling.  No internal level structure is required to accomplish this and cooling times on the order of seconds to minutes appear feasible for both axial and radial center-of-mass motion.  While we have not included the effects of absorption of light, which would heat the particle, early experiments by Ashkin suggest that microscale dielectric particles can be held for durations of 30 minutes before they are expelled from the field by radiometric forces \cite{ashkin}. These forces will be much less important in nanoscale dielectric particles since the particle will have a more uniform temperature. The effects of gas collisions must also be included in future work, as well as the finite size of the particle. The latter will slow the cooling process because the effective gradient in the axial potential will be smaller. Finally, these particles could also be cooled in the resolved-sideband limit, which for a fixed cavity geometry and reflectivity can be controlled by laser input intensity. In this regime cooling to the quantum limit is feasible and will be the goal of future work.

\end{document}